\documentclass[conference]{IEEEtran}
\IEEEoverridecommandlockouts
\usepackage[numbers,sort&compress]{natbib}
\usepackage{amsmath,amssymb,amsfonts}
\usepackage{graphicx}
\usepackage{textcomp}
\usepackage{amsmath}
\usepackage{booktabs} 
\usepackage{multirow}
\usepackage[colorlinks=true, linkcolor=blue, citecolor=blue, urlcolor=blue]{hyperref}

\usepackage{url}            
\usepackage{booktabs}       
\usepackage{amsfonts}       
\usepackage{nicefrac}       
\usepackage{microtype}      
\usepackage{xcolor}         
\usepackage{algorithm}
\usepackage{algpseudocode}
\usepackage{amsmath}
\usepackage{tikz}

\usepackage{orcidlink}
\usepackage{comment}
\usepackage{amssymb}
\usepackage{braket}
\usepackage{xcolor}
\def\BibTeX{{\rm B\kern-.05em{\sc i\kern-.025em b}\kern-.08em
    T\kern-.1667em\lower.7ex\hbox{E}\kern-.125emX}}

\usepackage{enumitem}

\usepackage{titlesec}
\titlespacing\section{0pt}{0.4\baselineskip}{0.2\baselineskip}
\titlespacing\subsection{0pt}{0.2\baselineskip}{0.1\baselineskip}
\titlespacing\subsubsection{0pt}{0.2\baselineskip}{0.1\baselineskip}

\begin{document}

\title{Quantum Clustering for Cybersecurity\vspace{-20pt}}

\author{\IEEEauthorblockN{Walid El Maouaki\textsuperscript{1}, Nouhaila Innan\textsuperscript{2,3}, Alberto Marchisio\textsuperscript{2,3}, Taoufik Said\textsuperscript{1}, Mohamed Bennai\textsuperscript{1}, Muhammad Shafique\textsuperscript{2,3}}
\IEEEauthorblockA{\textsuperscript{1}Quantum Physics and Magnetism Team, LPMC, Faculty of Sciences Ben M'sick,\\ Hassan II University of Casablanca,
Morocco\\
\textsuperscript{2}eBRAIN Lab, Division of Engineering, New York University Abu Dhabi (NYUAD), Abu Dhabi, UAE\\
\textsuperscript{3}Center for Quantum and Topological Systems (CQTS), NYUAD Research Institute, NYUAD, Abu Dhabi, UAE\\ walid.elmaouaki-etu@etu.univh2c.ma, nouhaila.innan@nyu.edu, alberto.marchisio@nyu.edu, taoufik.said@univh2c.ma, \\mohamed.bennai@univh2c.ma, muhammad.shafique@nyu.edu\\
}
\vspace{-30pt}
}

\maketitle

\begin{abstract}
In this study, we develop a novel quantum machine learning (QML) framework to analyze cybersecurity vulnerabilities using data from the 2022 CISA Known Exploited Vulnerabilities catalog, which includes detailed information on vulnerability types, severity levels, common vulnerability scoring system (CVSS) scores, and product specifics. Our framework preprocesses this data into a quantum-compatible format, enabling clustering analysis through our advanced quantum techniques, QCSWAPK-means and QkernelK-means. These quantum algorithms demonstrate superior performance compared to state-of-the-art classical clustering techniques like k-means and spectral clustering, achieving Silhouette scores of 0.491, Davies-Bouldin indices below 0.745, and Calinski-Harabasz scores exceeding 884, indicating more distinct and well-separated clusters. Our framework categorizes vulnerabilities into distinct groups, reflecting varying levels of risk severity: Cluster 0, primarily consisting of critical Microsoft-related vulnerabilities; Cluster 1, featuring medium severity vulnerabilities from various enterprise software vendors and network solutions; Cluster 2, with high severity vulnerabilities from Adobe, Cisco, and Google; and Cluster 3, encompassing vulnerabilities from Microsoft and Oracle with high to medium severity. These findings highlight the potential of QML to enhance the precision of vulnerability assessments and prioritization, advancing cybersecurity practices by enabling more strategic and proactive defense mechanisms.
\end{abstract}

\begin{IEEEkeywords}
Quantum Machine Learning, Quantum Clustering, Cybersecurity 
\end{IEEEkeywords}
\vspace{-10pt}

\section{Introduction}
The integration of quantum computing (QC) with cybersecurity has garnered significant attention, particularly with the increasing complexity and frequency of cyberattacks. Pioneered by Feynman and Manin, QC uses principles of quantum mechanics to tackle problems intractable for classical computers. Quantum effects like interference and entanglement enhance computational capabilities, making QC a target for cyber threats \cite{shara2023quantum}. As cyber threats become more sophisticated, protecting QC infrastructure from cyberattacks becomes crucial. 
Quantum machine learning (QML), which combines principles of QC with machine learning (ML) \cite{zaman2023survey}, has shown significant promise across various fields, including cybersecurity. Hybrid QML algorithms, which blend quantum and classical computations, have demonstrated effective applications in cybersecurity. For instance, a recent hybrid quantum-classical deep learning model successfully detected botnets \cite{suryotrisongko2022evaluating}. Additionally, QML can enhance intrusion detection systems by providing faster and more accurate solutions than classical methods, as evidenced in recent studies \cite{fioravanti2021evaluation,kalinin2023security}.  
By using quantum annealing, models can minimize free energy more efficiently than classical methods, resulting in better accuracy and faster training for cybersecurity tasks \cite{huanay2022applications}. These developments underscore the growing importance of QML in creating robust cybersecurity defenses.
Building on the advancements in QML, quantum clustering adapts traditional clustering algorithms to take advantage of QC's potential speedup capabilities. One approach employs variations of Grover's algorithm to quantize clustering methods like k-medians and neighborhood graph construction, achieving significant performance gains \cite{aimeur2007quantum}. Hybrid quantum k-means algorithms have shown theoretical and experimental benefits over classical clustering methods, especially for large datasets \cite{poggiali2024quantum}. Although traditionally used in modeling enzyme active sites, quantum chemical cluster approaches provide insights into how quantum clustering can be applied in various fields, including cybersecurity. These methods utilize quantum techniques to solve clustering problems more efficiently \cite{sheng2023quantum}. Recent studies have proposed quantum algorithms for anomaly detection in high-energy physics data, illustrating the versatility of quantum clustering in different domains \cite{wozniak2023quantum}. Quantum state clustering algorithms use variational quantum circuits to transform clustering into a parameter optimization problem, showing promising results in clustering quantum states \cite{fang2024quantum, bermejo2023variational}. The application of quantum paradigms in unsupervised ML, such as quantum k-means, further emphasizes the potential advantages of QC and QML in clustering tasks, particularly in handling large datasets more efficiently than classical methods \cite{kavitha2023quantum}.
While quantum clustering offers promising advancements, classical clustering algorithms remain vital in ML, especially cybersecurity. Unsupervised learning methods are used to identify false data injection attacks in smart grids, demonstrating the efficacy of clustering in detecting cyber threats \cite{pinto2023review}. Cluster analysis is also applied to categorize cybersecurity behaviors, allowing for tailored defense mechanisms specific to user profiles and enhancing overall security strategies \cite{baltuttis2024typology}. In vehicular networks, clustering algorithms manage communication and ensure data security and privacy, highlighting the importance of efficient clustering methods in dynamic environments \cite{kadam2023cybersecurity}. Predictive analytics and clustering are powerful tools in cybersecurity, enabling the identification of complex patterns and the automation of detection tasks. By clustering user behaviors and analyzing audit records, insider threats can be detected more effectively, showcasing the critical role of clustering in modern cybersecurity practices \cite{duary2024cybersecurity, nikiforova2024detecting}.
While QML has shown significant promise across various fields, \textit{applying quantum clustering specifically to cybersecurity remains unexplored.} The success of quantum clustering in other domains motivates its potential use in cybersecurity. Our study aims to bridge this gap by developing a quantum clustering framework to analyze cybersecurity vulnerabilities, thereby enhancing vulnerability assessments and prioritizing strategic defense mechanisms.
In this context, we focus on analyzing high-impact vulnerabilities from the 2022 CISA Known Exploited Vulnerabilities catalog, which includes diverse attack vectors and significant risks to various vendor systems and products. This targeted analysis demonstrates the applicability of quantum clustering to real-world security challenges and enhances our understanding of strategic vulnerability management.
\subsection{Contributions}
This study introduces two key advancements in the quantum K-means algorithm, an unsupervised QML technique. Our novel approach centers on developing two hybrid clustering methods that seamlessly integrate classical computing with quantum circuits. By emphasizing the strengths of both paradigms, we aim to enhance the algorithms' efficiency and scalability for complex datasets.
\begin{itemize}
    \item We introduce two methodologies for distance calculation in quantum K-means clustering: QCSWAPK-means and QkernelK-means. These algorithms iteratively update centroids and cluster assignments until they meet a predefined convergence criterion.
    \item 
We assess our quantum clustering methods, QCSWAPK-means and QkernelK-means, demonstrating significant performance improvements over the classical approaches across Silhouette, Davies-Bouldin, and Calinski-Harabasz scores.

    \item 
    We apply QCSWAPK-means and QkernelK-means to cluster the CISA dataset, revealing critical vulnerabilities and prioritizing urgent security concerns for strategic threat management.
\end{itemize}
\section{Framework}
\begin{figure}[htpb]
    \centering
    \includegraphics[width=\linewidth]{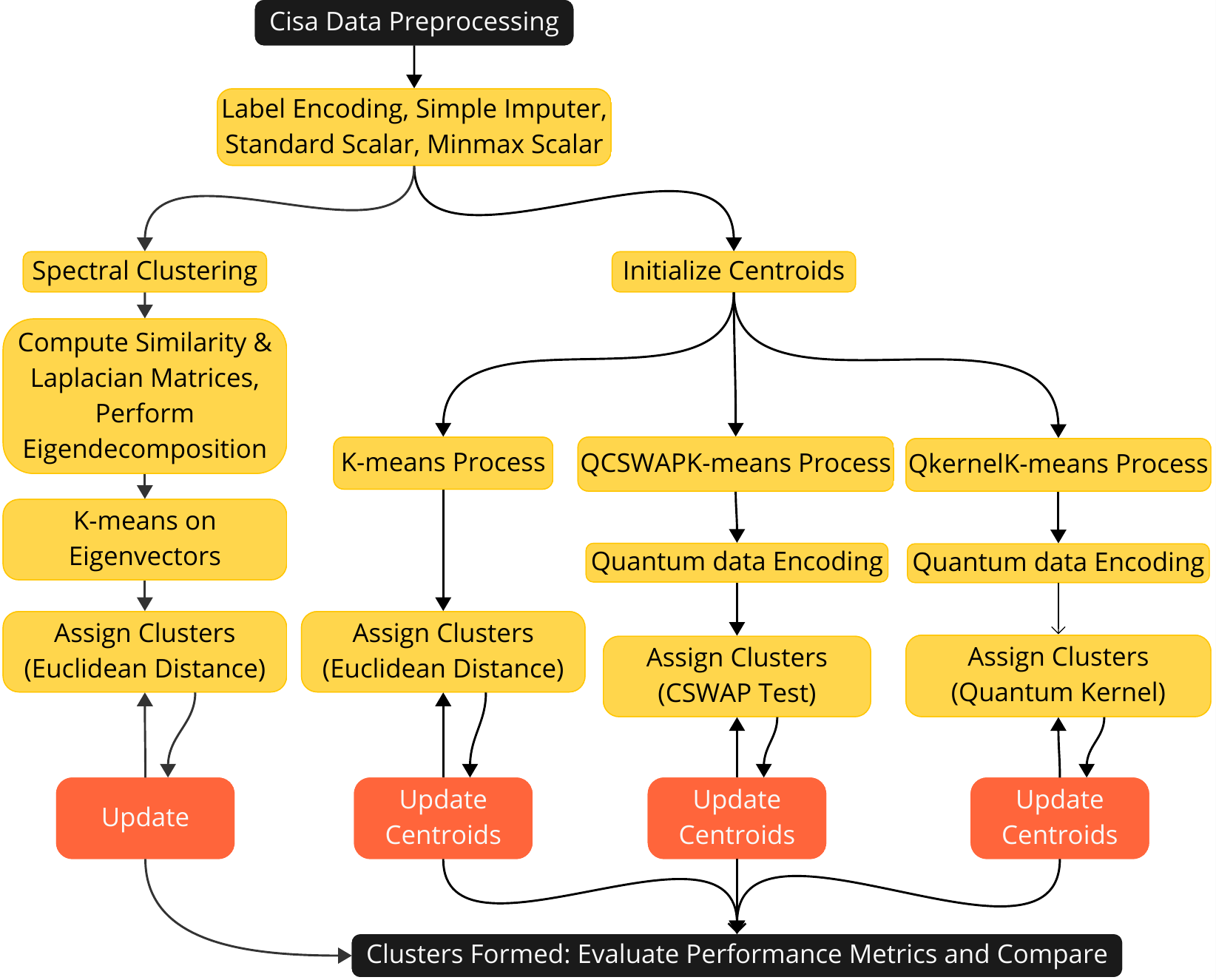}
    \vspace{-0.3cm}
    \caption{Clustering Process Workflow.}
    \label{Workflow}
\end{figure}

\subsection{Spectral clustering}
Spectral clustering is an advanced data analysis method that employs linear algebra to uncover groupings within datasets~\cite{von2007tutorial}. This approach starts by creating a graph representation of the data, where each point is a node, and their connections reflect their similarities. These similarities are typically computed using a Gaussian function.
The next step involves calculating the Laplacian matrix of this graph and analyzing its eigenvalues and eigenvectors. The algorithm focuses on the eigenvectors associated with the smallest eigenvalues, which contain crucial information about the data's structure. These selected eigenvectors are then combined to form a new, lower-dimensional representation of the original dataset.
In the final phase, a traditional clustering algorithm, such as K-means, is applied to this transformed data. By working in this modified space, spectral clustering can effectively identify clusters with complex shapes and structures that might be challenging for other methods to detect.

\subsection{Classical K-means}
The K-means algorithm is a widely used method for solving the unsupervised clustering problem \cite{likas2003global,sinaga2020unsupervised}. The algorithm aims to partition a dataset $X=\left\{\vec{x}_1, \ldots, \vec{x}_M\right\}$ of $M$ datapoints, where each datapoint $\vec{x}_i$ is an $N$-dimensional vector, into $k$ distinct clusters $\left\{C_1, \ldots, C_k\right\}$. Each cluster is represented by a centroid $\left\{\vec{c}_1, \ldots, \vec{c}_k\right\}$, such that similar records are grouped together based on a specific distance measure. The K-means algorithm operates iteratively through two primary steps until convergence. Initially, $k$ centroids are selected randomly from the dataset. The first step is the cluster assignment step, where each datapoint in the dataset is assigned to the nearest centroid. This is typically done using the Euclidean distance, defined as:
\begin{equation}
    D(\vec{a}, \vec{b})=\|\vec{a}-\vec{b}\|_2=\sqrt{\sum_{i=1}^N\left(a_i-b_i\right)^2},
\end{equation}
where $\vec{a}$ and $\vec{b}$ are $N$-dimensional vectors. The second step involves updating the centroids by computing the mean of all records assigned to each cluster. These updated centroids are then used in the next iteration, see Fig.~\ref{Workflow} for an overview of the process.
The objective of K-means is to minimize the within sum of squared errors (WSSE), which is given by:
\begin{equation}
E=\sum_{k=1}^K \sum_{m \in C_k}\left\|x_m-c_k\right\|^2,
\end{equation}
where $x_n$ represents the datapoints and $c_k$ represents the centroids. One method for determining the optimal number of clusters, $k$, is the ``elbow method''. This involves plotting the within-cluster sum of squares (WCSS) against the number of clusters and looking for an ``elbow'' point where the rate of decrease sharply changes~\cite{shi2021quantitative}. In terms of computational complexity, the classical K-means algorithm has a time complexity of $O(kMN)$, where $k$ is the number of clusters, $M$ is the number of records, and $N$ is the dimensionality of the data. This complexity can become a limiting factor for very large datasets, leading researchers to explore more efficient alternatives, such as quantum-based approaches.

\subsection{Quantum K-means}
To leverage quantum-inspired approaches for distance computation in K-means clustering, it is crucial to first map classical data into quantum states through quantum data encoding. Various encoding methods exist, but two particularly relevant techniques are quantum amplitude encoding and angle encoding. In quantum amplitude encoding, a classical data vector $\vec{x}= ( x_1, x_2, \dots, x_N )$ is mapped into the amplitudes of a quantum state $| \psi \rangle = \sum_{i=1}^N x_i |i \rangle$, where $|i\rangle$ represents the computational basis states. This encoding typically requires $\log_2(N)$ qubits to represent $N$ data points, making it efficient for large datasets. Angle encoding, on the other hand, maps each classical data point to a qubit's state on the Bloch sphere as $|\psi\rangle =\cos(x) |0\rangle +\sin(x) |1\rangle$, this method requires one qubit per feature of the data vector. This encoding is particularly useful for data with values in the range $[0, \pi]$. By encoding classical data into quantum states, we can exploit quantum parallelism and interference to potentially accelerate distance calculations in K-means, opening up new avenues for efficient clustering of high-dimensional datasets. \textit{The results in this paper are produced using the angle encoding technique.}

\subsubsection{\textbf{QCSWAPK-means}}
We introduce QCSWAPK-means, which employs the Controlled SWAP (CSWAP) test, initially described in~\cite{buhrman2001quantum}, to calculate the distance between quantum states. Our approach uses the CSWAP test to enhance the cluster assignment phase of the k-means algorithm, enabling more accurate and efficient data clustering. The CSWAP test involves a quantum circuit with an ancillary qubit and additional qubits encoding the quantum states of the corresponding datapoints. The process begins by applying a Hadamard gate to the ancilla qubit, preparing it in the superposition state $\frac{1}{\sqrt{2}}(|0\rangle+|1\rangle)$. Subsequently, two quantum states representing the datapoint $|\psi(\vec{x})\rangle$ and the centroid $|\psi(\vec{c})\rangle$ are loaded into separate sets of qubits. The state of the quantum circuit at this point is
$\frac{1}{\sqrt{2}}(|0\rangle \otimes|\psi(\vec{x})\rangle \otimes|\psi(\vec{c})\rangle+|1\rangle \otimes|\psi(\vec{x})\rangle \otimes|\psi(\vec{
c})\rangle)$. Next, a CSWAP gate, conditioned on the state of the ancilla qubit, is applied to the qubits encoding $|\psi(\vec{x})\rangle$ and $|\psi(\vec{c})\rangle$. This operation transforms the state into
$
\frac{1}{\sqrt{2}}(|0\rangle \otimes|\psi(\vec{x})\rangle \otimes|\psi(\vec{c})\rangle+|1\rangle \otimes|\psi(\vec{c})\rangle \otimes|\psi(\vec{x})\rangle)$. Another Hadamard gate is then applied to the ancilla qubit, evolving the state into:
$
\frac{1}{2}(|0\rangle \otimes(|\psi(\vec{x})\rangle \otimes|\psi(\vec{c})\rangle+|\psi(\vec{c})\rangle \otimes|\psi(\vec{x})\rangle)+|1\rangle \otimes(|\psi(\vec{x})\rangle \otimes|\psi(\vec{c})\rangle-|\psi(\vec{c})\rangle \otimes|\psi(\vec{x})\rangle))
$. The final step involves measuring the ancilla qubit. The probability of measuring the ancilla in the state $|0\rangle$ is given by $P(|0\rangle)=\frac{1+|\langle \psi(\vec{x}) \mid \psi(\vec{c})\rangle|^2}{2}$. The overlap $|\langle \psi(\vec{x}) \mid \psi(\vec{c}) \rangle|$ quantifies the similarity between the two states, thereby providing a measure of the distance between the corresponding datapoints. The assignment of datapoints to clusters is determined by this overlap, facilitating the clustering process with quantum-enhanced computation.

\subsubsection{\textbf{QkernelK-means}}
Inspired by the quantum support vector machine (QSVM) algorithm~\cite{rebentrost2014quantum}, we introduce the QkernelK-means method, which utilizes quantum kernels to calculate the overlap between data points $\vec{x}$ and centroids $\vec{c}$. This is achieved through encoding the data into quantum states via unitary transformations.
To begin, a data point $\vec{x}$ is mapped to a quantum state using a unitary operation $U(\vec{x})$, resulting in $|\psi(\vec{x})\rangle=U(\vec{x})|0\rangle$. Similarly, the centroid $\vec{c}$ is encoded as $|\psi(\vec{c})\rangle$ using $U(\vec{c})$. The overlap, or similarity, between these states is computed by applying the adjoint unitary operation $U^{\dagger}(\vec{c})$ to $|\psi(\vec{x})\rangle$, yielding:
$
K(\vec{x}, \vec{c})=|\langle\psi(\vec{c}) \mid \psi(\vec{x})\rangle|^2=\left|\left\langle 0\left|U^{\dagger}(\vec{c}) U(\vec{x})\right| 0\right\rangle\right|^2
$.
This overlap provides a measure of similarity between the data point and the centroid. The probability of measuring the resulting quantum state in the ground state $|0\rangle$ gives the quantum kernel value, which assigns datapoints to their respective clusters based on their similarity. The QkernelK-means method provides a powerful tool for clustering in high-dimensional spaces, enhancing computational efficiency and improving the clustering performance by exploiting the quantum properties of the data.
Refer to Fig.~\ref{Workflow} for a comprehensive overview of the clustering process workflow, comparing the four algorithms.

\section{Results and Discussion}
\subsection{Experimental Setup and Results}
\begin{figure}[h]
    \centering
       \vspace{-0.2cm}
    \includegraphics[width=0.7\linewidth]{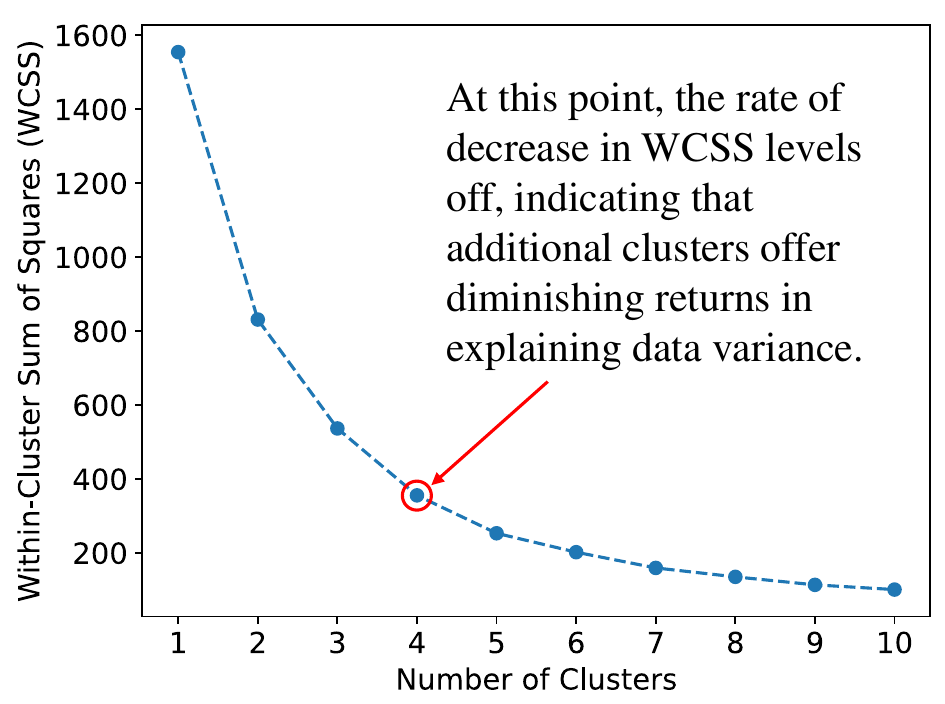}
    \vspace{-0.2cm}
    \caption{The elbow technique identifies the optimal number of clusters for K-means, showing the number of clusters (k) versus the WCSS. The ``elbow'' of the curve, indicating the optimal number of clusters, occurs at k=4.} 
    \label{elbow}
\end{figure}

The dataset used in this study is derived from the CISA Known Exploited Vulnerabilities catalog for 2022 \cite{dataset}, encompassing detailed information on various security vulnerabilities. It includes attributes such as CVE ID, vendor project, product name, vulnerability name, date added, short description, required action, due date, common vulnerability scoring system (CVSS) score, CWE, attack vector, complexity, and severity. 
Our experiments include an analysis of vulnerability patterns across different vendors and products. We employ clustering techniques to uncover groups of vendors and products that exhibited similar vulnerability profiles. To initiate this process, we focus on two key attributes from our dataset: the vendor or project name and the specific product identifier. These columns are chosen as they provide crucial information about the origin and nature of each vulnerability, as well as uncover relationships and potential dependencies within the software ecosystem. This information can be valuable for understanding how vulnerabilities propagate across different vendors and products. While features like complexity and severity are critical for assessing individual vulnerabilities, our clustering aims to reveal broader patterns in the vulnerability landscape at the vendor and product levels. The clustering analysis is done using four different algorithms: Spectral clustering, K-means, QCSWAPK-means, and QkernelK-means. The optimal number of clusters is determined to be four based on the elbow method applied to the K-means algorithm, see Fig.~\ref{elbow}.

The vendor/project and product features represent categorical data. To make these features compatible with machine learning algorithms and quantum computing methods, we employed the label encoding technique to map categorical data into numerical data. This process assigns a unique integer to each category within the features. Consequently, each vendor and product is labeled by a distinct numerical value, resulting in two numerical features for our dataset. Our final dataset comprises 777 data samples. Given that our preprocessed data now has two numerical features, we utilize the angle encoding technique to represent this classical data in quantum states. Angle encoding requires one qubit per feature to encode the data into a quantum state. Therefore, for our two-feature dataset, we need two qubits to fully represent each data point in the quantum realm.

\begin{figure}[htpb]
    \centering
    \includegraphics[width=\linewidth]{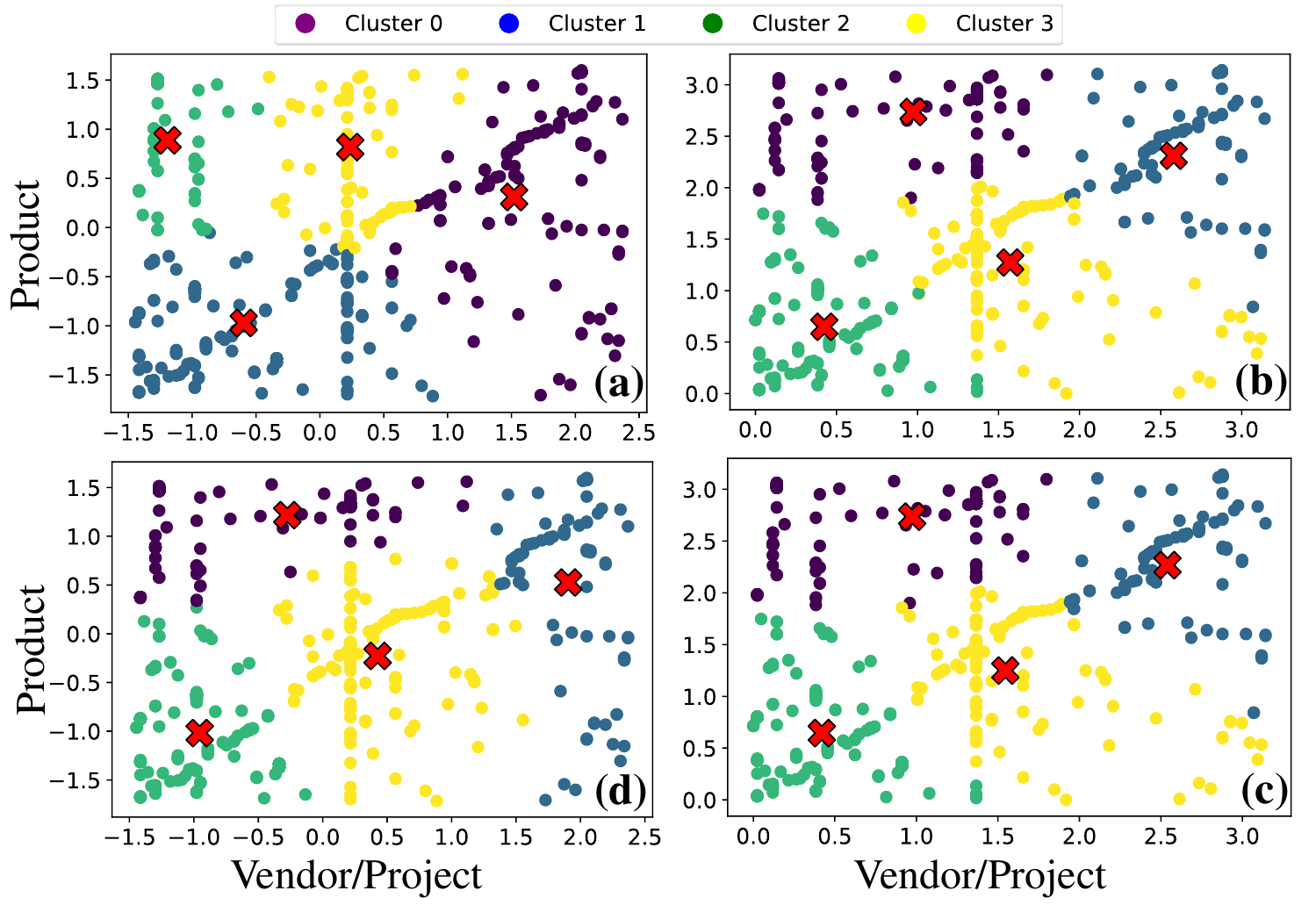}
    \vspace{-0.6cm}
    \caption{Clustering outcomes using four algorithms: a) K-means, b) QCSWAPK-means, c) QkernelK-means, d) Spectral clustering. Subplots show clusters in 2D space (Vendor/Project vs. Product). Red X markers indicate centroids. In the classical algorithms, the dataset was normalized to fall within the range of $[0, 1]$. The quantum-enhanced methods QCSWAPK-means and QkernelK-means are normalized to $[0, \pi]$ to meet the specific requirements of quantum encoding.}
    \label{scatter}
\end{figure}

For the QCSWAP-Kmeans algorithm implementation, we require a total of five qubits: two qubits to encode the centroid data point, two qubits to encode the data point we want to cluster, and one additional qubit to measure the similarity between the data points. In contrast, for the Qkernel-Kmeans algorithm, we encode the data points' states in parallel, which allows us to use only two qubits in total. In this case, we measure the first qubit to determine the similarity between data points.
The experiments are conducted on a quantum simulator using PennyLane.
The scatter plots in Fig.~\ref{scatter} are generated for each algorithm and visually demonstrate the formation of four distinct clusters, with X markers highlighting the centroid points of the clusters. The Spectral clustering and K-means algorithms showed a relatively clear separation of clusters but exhibited a different clustering configuration compared to the QCSWAPK-means and QkernelK-means algorithms. In contrast, the QCSWAPK-means and QkernelK-means algorithms produced more refined and compact clusters, indicating enhanced clustering performance.
The effectiveness of the clustering algorithms is assessed using the three evaluation metrics: the Silhouette score, the Davies-Bouldin index, and the Calinski-Harabasz index, as presented in Table \ref{clustering_comparison}. The results reveal that the QCSWAPK-means algorithm achieves the highest Silhouette score of 0.491, suggesting more distinctly defined clusters compared to the K-means and Spectral clustering algorithms, which score 0.451 and 0.486, respectively. Additionally, the QCSWAPK-means algorithm demonstrates superior clustering quality with a Davies-Bouldin index of 0.739 and a Calinski-Harabasz index of 884.594. The QkernelK-means algorithm also shows competitive performance, recording a Silhouette score of 0.491, a Davies-Bouldin index of 0.745, and a Calinski-Harabasz index of 885.954. These findings substantiate the enhanced effectiveness of the QCSWAPK-means and QkernelK-means algorithms over the conventional K-means and Spectral clustering methods, which exhibit a different clustering configuration, resulting in lower performance metrics.
\begin{table}[h]
\centering
\vspace{-0.1cm}
\caption{Clustering Algorithms Performance Comparison.}
\vspace{-0.1cm}
\label{clustering_comparison}
\resizebox{\columnwidth}{!}{%
\begin{tabular}{lccc}
\hline
\textbf{Algorithm} & \textbf{Silhouette $\uparrow$} & \textbf{Davies-Bouldin $\downarrow$} & \textbf{Calinski-Harabasz $\uparrow$} \\
\hline
K-means & 0.451 & 0.883 & 678.137 \\
Spectral Clustering & 0.486 & 0.785 & 799.770 \\
CSWAPK-means & \textbf{0.491} & \textbf{0.739} & 884.594 \\
QkernelK-means & \textbf{0.491} & 0.745& \textbf{885.954} \\
\hline
\end{tabular}%
}
\end{table}
The clustering results from the classical algorithms (K-means and Spectral clustering) and the quantum algorithms (CSWAPK-means and QkernelK-means) reveal distinct patterns in the distribution of vulnerabilities among various vendors and products. The K-means algorithm identifies four clusters with specific vendor focus areas: Cluster 0 groups diverse enterprise software and network solutions without a single dominating vendor; Cluster 1 is dominated by major tech and security giants such as Microsoft, Adobe, Cisco, and Google; Cluster 2 focuses significantly on Apple and other major enterprise software vendors; and Cluster 3 is heavily dominated by Microsoft. The Spectral clustering algorithm identifies four clusters with distinct vendor emphases: Cluster 0 groups various technology solutions with Microsoft as the leader, supplemented by Oracle, Mozilla, and other tech vendors. Cluster 1 focuses on security-related firms like VMware and SonicWall, among others with no single vendor dominating. Cluster 2 combines major tech firms such as Microsoft and Apple with other significant contributors like Apache and Cisco. Cluster 3 is dominated by Adobe and Cisco, with strong participation from Google.

In contrast, the quantum algorithms yield a slightly different clustering configuration, each reflecting varying levels of risk severity. Cluster 0, characterized by a significant concentration of vulnerabilities related to Microsoft products, reflects critical severity due to its extensive use and broad risk exposure. Cluster 1 features a mix of enterprise software vendors and network solutions, exhibiting medium severity, similar to K-means Cluster 0. Cluster 2, notable for high-severity vulnerabilities concentrated in Adobe, Cisco, and Google products, underscores their susceptibility to a wide range of severe exploits. Finally, Cluster 3, which focuses on vulnerabilities in Microsoft and Oracle products, indicates a severity range from high to medium, highlighting the critical need for robust security measures for these widely used platforms.

\subsection{Discussion}
The results of the clustering experiments provide insights into the performance of the K-means and Spectral clustering algorithms compared to the novel QCSWAPK-means and QkernelK-means algorithms. The classical algorithms, while effective at identifying distinct clusters, show limitations in the compactness and separation of the clusters, as reflected in its Silhouette score of 0.451 for K-means and 0.486 for Spectral clustering. This indicates that data points might not be well-separated from other clusters, leading to some ambiguity in cluster assignment. Additionally, the Davies-Bouldin index of 0.883 and the Calinski-Harabasz index of 678.137 for K-means suggest that while the clusters are reasonably distinct, there is room for improvement in terms of cluster cohesion and separation. The different clustering configurations exhibited by K-means compared to QCSWAPK-means and QkernelK-means likely contribute to its lower clustering performance.
In contrast, both QCSWAPK-means and QkernelK-means achieve higher Silhouette scores of 0.491, suggesting better-defined clusters. Additionally, their lower Davies-Bouldin indices (0.739 and 0.745, respectively) and higher Calinski-Harabasz indices (884.594 and 885.954, respectively) demonstrate superior cluster cohesion and separation. These results validate the enhanced effectiveness of the proposed algorithms in handling the complex patterns within the cybersecurity dataset, producing more accurate and meaningful clusters compared to the traditional K-means algorithm.


Quantum-inspired clustering effectively identifies critical security areas by emphasizing concentrations of vulnerabilities, particularly within certain vendors. For instance, Microsoft's presence in multiple clusters (0 and 3) indicates a larger attack surface due to its diverse product range. Similarly, moderate vulnerability counts for Apple, Apache, and Cisco in Cluster 0, and for Oracle in Cluster 3, suggest potential common security weaknesses or a higher frequency of targeting. This method highlights the potential of quantum algorithms to provide a nuanced understanding of vendor-specific vulnerabilities, directing resources towards vendors with higher vulnerability counts to potentially enhance the overall security posture.

\section{Conclusion}
In this study, we propose two quantum clustering approaches to analyze and categorize cybersecurity vulnerabilities. By processing data from the 2022 CISA Known Exploited Vulnerabilities catalog, we demonstrate the effectiveness of our quantum clustering approach in identifying distinct clusters of vulnerabilities, each reflecting varying levels of risk severity. Our results reveal that quantum clustering methods, specifically our QCSWAPK-means and QkernelK-means algorithms, produce more refined and compact clusters compared to traditional K-means and Spectral clustering, indicating enhanced clustering performance. The quantum algorithms' ability to better define clusters and improve separation metrics highlights their potential in advancing vulnerability assessments and prioritizing strategic cybersecurity defenses.
Our findings underscore the significant potential of QML in enhancing cybersecurity practices by providing a more nuanced understanding of vulnerability patterns. The distinct clusters identified by the quantum algorithms not only pinpoint critical areas with high-risk vulnerabilities but also suggest strategic directions for resource allocation and defense prioritization. Future research should explore the impact of quantum circuit specifics, like data encoding, on the results and experiment with larger datasets. This work paves the way for integrating QC into cybersecurity, offering a promising avenue for more proactive and strategic defense mechanisms.
\section*{Acknowledgment}
This work was supported in parts by the NYUAD Center for Quantum and Topological Systems (CQTS), funded by Tamkeen under the NYUAD Research Institute grant CG008, and the NYUAD Center for Cyber Security (CCS), funded by Tamkeen under the NYUAD Research Institute Award G1104.

\bibliographystyle{IEEEtran}

\bibliography{refs}

\end{document}